\documentclass[aps,amsfonts,amsmath,prd,preprint,nofootinbib]{revtex4}
\usepackage[pdftex]{graphicx}
\usepackage{enumerate}
\usepackage{amsthm}

\newcommand{\beq}{\begin{equation}}
\newcommand{\eeq}{\end{equation}}

\begin{document}

\title{Cosmological singularity theorems and black holes}

\author{Alexander Vilenkin$^1$ and Aron C. Wall$^2$}
\address{
$^1$ Institute of Cosmology, Department of Physics and Astronomy,\\
Tufts University, Medford, MA 02155, USA\\
$^2$ Department of Physics, University of California,\\
Santa Barbara, California 93106, USA}

\begin{abstract}

An extension of Penrose's singularity theorem is proved for spacetimes where black holes are allowed to form from non-singular initial data.  With standard assumptions about the spacetime, and assuming the existence of a trapped surface which lies outside of black hole horizons and is not completely surrounded by horizons, we show that the spacetime region outside (or on) the horizons must contain singularities.  If the trapped surface is surrounded by horizons, we show that the horizons divide spacetime into causally disconnected pieces.  Unlike the original Penrose's theorem, our theorems provide some information about the location of singularities.  We illustrate how they can be used to rule out some cosmological scenarios.

\end{abstract}

\maketitle

\section{Introduction}

The singularity theorems of General Relativity demonstrate that spacetimes having certain properties must necessarily be singular.  The assumptions of the theorems are rather general and mild; in this sense these are very powerful results.  Their main weakness is that they do not say much about the nature of the singularities.

Consider for example the question that motivated some of the early work on the subject.  It has been known since Friedmann and Lemaitre that contracting open homogeneous and isotropic (FLRW) models collapse to a big crunch.  But could it be that the singularity is an artifact of the symmetry of these models?
If the contracting universe is made up of clumps, could the clumps miss one another and fly apart, so that the contraction is followed by an expansion?  This picture was championed by Khalatnikov and Lifshitz (KL) \cite{Lifshitz}, who argued that the most general solution
of Einstein's equations would describe a non-singular contracting and re-expanding universe.  Penrose's singularity theorem \cite{Penrose} (see also Hawking \cite{Hawking}) dealt a fatal blow to this idea.  The key new concept introduced by Penrose was that of a trapped surface---a compact, 2-dimensional\footnote{More generally, in a $D$ dimensional spacetime, a trapped surface will be $D - 2$ dimensional.} surface ${\cal T}$ such that both outward and inward directed systems of null geodesics emanating orthogonally from ${\cal T}$ toward the future are converging at all points on ${\cal T}$.  The existence of such a surface, combined with mild assumptions about the global structure of spacetime and about its matter content (the null energy condition), makes singularities in the future of ${\cal T}$ unavoidable.  More precisely, at least one future-directed null geodesic emanating orthogonally from ${\cal T}$ must terminate at a singularity.  Trapped surfaces do exist in all contracting open FLRW models (any sphere of sufficiently large radius is a trapped surface), and it appears that such surfaces must exist even if the assumption of homogeneity and isotropy is lifted and matter is distributed in clumps.  Thus, barring nontrivial global properties or exotic matter content, a contracting infinite universe must be singular and the KL picture cannot be right.

A time-reversed version of this argument can be applied to an expanding infinite universe.  Such a universe necessarily contains an anti-trapped surface on which all past-directed null geodesics are converging.  Then at least one of these geodesics must terminate at a singularity in the past.

On the other hand, the Penrose theorem tells us very little about the location or the nature of the singularities.  Suppose, for example, that the KL picture of a clumpy contracting and re-expanding universe
is basically correct, except some of the clumps merge in the course of contraction and turn into black holes.  
Some future-directed null geodesics will then enter black holes and terminate at singularities, so this modified KL scenario is consistent with the theorem and cannot be excluded.

\begin{figure}[ht]
\begin{center}
\includegraphics[width=11cm]{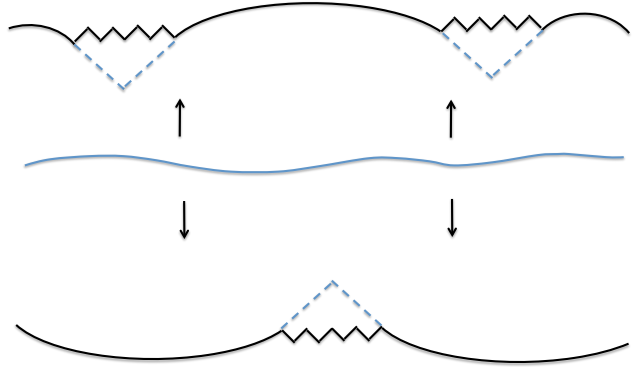}
\caption{Spacetime diagram representing the spontaneous inflation scenario.  The Cauchy surface is shown by a blue curved line.  Black hole (and time-reversed black hole) singularities, shown by zigzag lines, are hidden behind horizons (dashed lines).  The arcs represent spacelike future and past infinities of the inflating regions.   Thermodynamic arrows of time are indicated by arrows.}
\label{carroll}
\end{center}
\end{figure}

Another interesting example is the spontaneous inflation scenario of Carroll and Chen \cite{CarrollChen} (CC), which  
starts with a typical\footnote{CC use the word ``generic'', but since this word also has some more precise meanings in general relativity proofs \cite{HawkingEllis, Wald}, we substitute the word typical.} initial state on an infinite spacelike Cauchy surface ${\cal S}$.\footnote{The assumption that the Cauchy surface is infinite is stated on p.19 of Ref.~\cite{CarrollChen}.}  Assuming a stable vacuum with a positive energy density, it is then argued that the resulting spacetime will approach a vacuum de Sitter state in both time directions.  The universe will be contracting in the asymptotic past and expanding in the asymptotic future, with the thermodynamic arrow of time pointing away from ${\cal S}$.   
The asymptotic de Sitter regions in the past (future) of ${\cal S}$ contain trapped (anti-trapped) surfaces, and the singularity theorem requires that some future and past-directed null geodesics must be incomplete.  But once again a contradiction with the theorem is avoided if some black holes are formed in the future of ${\cal S}$ and some time-reversed black holes (i.e. white holes) are formed in the past of ${\cal S}$.  The resulting spacetime structure is illustrated in Fig.~\ref{carroll}.

The purpose of the present paper is to extend the Penrose singularity theorem to cosmological spacetimes where black holes are allowed to form from non-singular initial data.  Assuming the cosmic censorship hypothesis \cite{Penrose2}, for generic initial data, all singularities are hidden behind event horizons.  There is also a stronger version of the hypothesis \cite{Penrose3} which says that singularities cannot even lie on the horizon; this rules out ``thunderbolt''-type situations where a null singularity expands outward at the speed of light.  For our purposes, we will consider all of these to be ``naked'' singularities; a more precise definition will be given in section \ref{inc}.

\begin{figure}[htp] 
\begin{center}
\includegraphics[width=11cm]{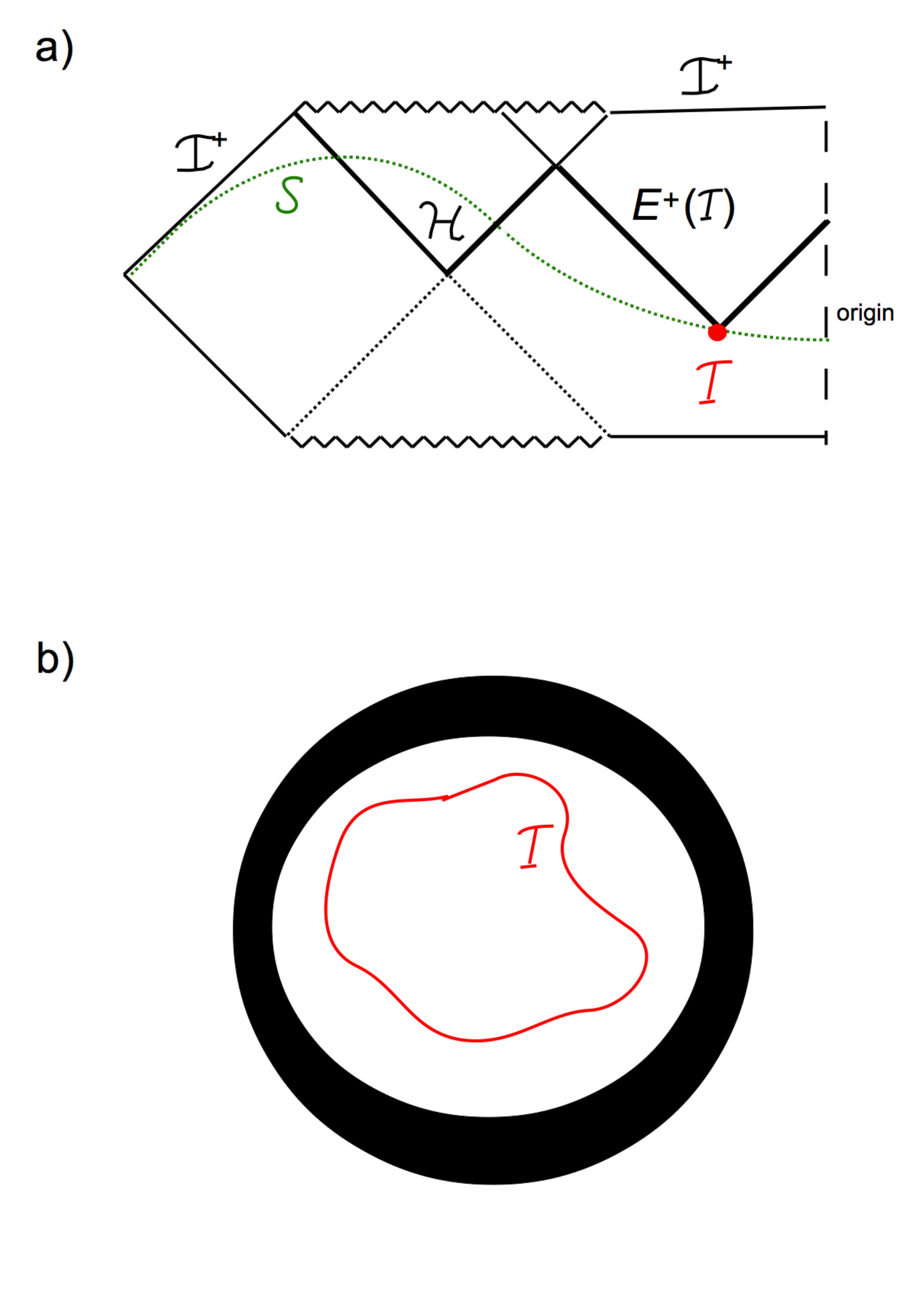}
\caption{(a) A causal diagram of the ``bag of gold'' spacetime, where asymptotically de Sitter and Minkowski regions are connected by an Einstein-Rosen bridge. The green dotted line indicates a spacetime slice ${\cal S}$, which is a Cauchy surface for this spacetime.  The black hole horizon ${\cal H}$ and the future light cone of the trapped surface ${\cal T}$ are joined to form the surface ${\cal E}$, which is shown by a thick black line.  (b) The trapped surface ${\cal T}$ as it appears in a spacelike slice ${\cal S}$ containing ${\cal T}$.  The black ring represents the black hole, with its inner and outer boundaries representing horizon components on the de Sitter and Minkowski sides, respectively.  The surface ${\cal T}$ is completely surrounded by the black hole horizon.}
\label{surround}
\end{center}
\end{figure}

We shall assume that black hole horizons are sufficiently sparse that they do not completely surround the trapped surface ${\cal T}$.\footnote{For asymptotically flat spacetimes which are also ``asymptotically predictable'', trapped surfaces ${\cal T}$ are required to lie completely inside of a black hole horizon \cite{HawkingEllis, Wald}.  However, this theorem does not necessarily apply to cosmological settings.} 
As an example of a situation where ${\cal T}$ is completely surrounded by the horizon, consider the ``bag of gold" spacetime \cite{deWitt}, where a contracting and re-expanding asymptotically de Sitter space is connected to an asymptotically Minkowski space by an Einstein-Rosen bridge.  The causal diagram for this spacetime is shown in Fig.~\ref{surround}(a).  The spacelike surface ${\cal S}$ indicated by a dotted green line is a Cauchy surface for this spacetime.  This surface has the geometry of a plane which is connected by a neck to a sphere.  As this initial data is evolved either to the past or to the future, a black hole is formed in the neck region.  The horizon bounding this black hole has two components, facing the de Sitter and Minkowski sides, respectively.  The black hole thus has the topology of a hollow sphere (a black 2-brane).  Now, the contracting part of de Sitter contains trapped surfaces; one such surface is indicated by a red dot in the figure.   This surface ${\cal T}$, which is shown in Fig.~\ref{surround}(b) as it appears in a spacelike slice ${\cal S}$ containing ${\cal T}$, is completely surrounded by the horizon.

On the other hand, a set of isolated compact black holes, each with a connected horizon, cannot completely surround a surface.  One could also imagine a more complicated ``black system'' consisting of e.g. a network of black strings, which would still not completely surround ${\cal T}$ if there are ``holes'' through which a timelike worldline could escape without falling across the horizon.  In situations like these, we will show that there must be \emph{additional} (i.e. naked) singularities besides those which are inside of the black holes.  If we stipulate that there are no naked singularities, this is a contradiction.  Thus we are forced to drop one of the other assumptions, concluding that either (i) the universe is actually spatially compact, or (ii) the singularities are so extensive that each trapped surface ${\cal T}$ is either inside, or else completely surrounded by, an event horizon.

We shall also prove a second version of this theorem, where instead of assuming that ${\cal T}$ is not surrounded by the horizon, we require that all black holes originate at some finite time and have horizons with just one connected component.  In this case we again get a contradiction with the existence of a trapped surface.

These theorems place restrictions on the KL and CC scenarios, even when modified to include isolated black hole singularities.  They show that the singularities appearing in a spatially noncompact universe must be of a more ``cosmological'' character than black hole singularities.  We cannot show that every causal curve encounters a singularity---a counterexample is the ``bag of gold" spacetime of Fig.~\ref{surround}. 
However, if you start inside a trapped surface, then successful evasion of the singularity requires entering a compact bubble universe which is completely surrounded by event horizons.

In the next Section, we briefly outline the proof of the standard Penrose singularity theorem; the proofs of our theorems follow the same pattern.  Our extended theorems are formulated and proved in Section III.  Finally, our results are summarized and discussed in Section IV.

We use standard notation in which $I^\pm(X)$ is the chronological future/past of a set $X$, $J^\pm(X)$ is the causal future/past of $X$ (which also includes lightlike separated points), and $\partial X$ is the boundary of $X$.

\section{The Penrose theorem}\label{pen}

In this Section we shall outline the standard proof of the Penrose theorem.
A variety of singularity theorems have been proved in the literature, but the number of basic ideas and methods of proof is rather limited.  For a comprehensive review of the subject, see Refs.~\cite{HawkingEllis,Wald}.  The Penrose theorem says that spacetime which (a) has a non-compact, connected Cauchy surface ${\cal S}$, (b) obeys the null convergence condition, and (c) contains a trapped surface ${\cal T}$ must be null incomplete to the future.  Specifically, at least one of the future-directed null geodesics emanating orthogonally from ${\cal T}$ must be incomplete.

The existence of a Cauchy surface implies a rather simple causal structure of spacetime, namely that spacetime is globally hyperbolic.\footnote{Globally hyperbolic spacetimes are those for which hyperbolic wave equations have well-defined initial value problems.  They satisfy two requirements: (i) there are no closed causal curves, and (ii) for any two points $p$ and $q$, the region $J^+(p) \cap J^-(q)$ is compact.  (Traditionally, the definition of global hyperbolicity included a stronger assumption than (i), known as ``strong causality'', but these two definitions of global hyperbolicity have been proven to be equivalent \cite{BernalSanchez}.)  Condition (i) forbids time travel paradoxes, and condition (ii) ensures that no information leaks in or out of the spacetime.  In particular, it rules out ``timelike'' singularities, which are forbidden by certain versions of the cosmic censorship hypothesis \cite{Penrose3}.}
Here, we are not going to be interested in causality violations, so this is a reasonable assumption to adopt.  

The null convergence condition (NCC) requires that $R_{\mu\nu}N^\mu N^\nu \geq 0$ for all null vectors $N^\mu$, where $R_{\mu\nu}$ is the Ricci tensor.  
Combined with Einstein's equations, NCC is equivalent to the null energy condition (NEC), requiring that $T_{\mu\nu}N^\mu N^\nu \geq 0$ for all null $N^\mu$, where $T_{\mu\nu}$ is the energy-momentum tensor.  
NEC is normally satisfied for reasonable classical matter models that are minimally coupled to the metric.
It can be violated by quantum effects, but such violations are localized and restricted by "quantum inequalities" (see, e.g. Ref.~\cite{Ford} and references therein).\footnote{Many of the results based on NCC can also be derived using weaker conditions, such as the integral convergence condition \cite{Tipler,Roman}, repeated integral convergence condition \cite{Borde}, or the generalized second law \cite{Wall}.}  We shall disregard possible violations of NCC in what follows.

Although the Penrose theorem is normally regarded as a ``singularity theorem'', by focusing on premise (a) one can instead cast it as a theorem which requires the universe to be spatially compact given certain assumptions.  If one assumes that our universe was singularity-free to the past, then the time-reverse of the ``open universe'' theorem implies that the universe must in fact be spatially compact.  Note that if a cosmology admits a spatially compact but nonsimply connected Cauchy surface ${\cal S}$, we can move to the universal cover of ${\cal S}$ to apply the theorem \cite{Tipler}.  Thus the theorem places restrictions on open or flat cosmologies, even if the universe has nontrivial topology.  However, it does not violate the theorem to include a seemingly open universe inside of an expanding bubble within a spatially closed universe.

The standard proof of the Penrose theorem is by contradiction:
\begin{proof}
Suppose the conditions (a)-(c) are satisfied and the spacetime is null complete to the future.  Let ${\cal T}$ be a trapped surface.  The future light cone of ${\cal T}$, $E^+({\cal T})$, can be defined as the boundary of its future, 
\beq
E^+({\cal T})={\partial I^+}({\cal T}). 
\label{E+}
\eeq
In the vicinity of ${\cal T}$, the light cone is comprised of the two future-directed sheets of null geodesics emanating from ${\cal T}$.  The null geodesics on $E^+({\cal T})$ converge, and it follows from NCC and future null completeness that each of these geodesics comes to a conjugate point (that is, crosses nearby geodesics on $E^+({\cal T})$) in a finite affine parameter time.  After crossing, the geodesics do not stay on the light cone and enter its interior.  Since this happens to all geodesics in a finite affine time, the light cone must be compact.  And since $E^+({\cal T})$ is a boundary of a set $(I^+({\cal T}))$, it must have no boundary.  It must also be achronal, which means that no two points on $E^+({\cal T})$ can be connected by a timelike curve.  (Otherwise, one of these points would be in the interior of $I^+({\cal T})$.) 

Now, the existence of a compact, edgeless, achronal hypersurface is inconsistent with a non-compact, connected Cauchy surface.  In order to see this, consider a smooth timelike vector field $V^\mu(x)$ whose integral curves cross the Cauchy surface ${\cal S}$ exactly once.  (The existence of such a field follows from the fact that ${\cal S}$ is a Cauchy surface.)  Since the light cone $E^+({\cal T})$ is achronal, the integral curves of $V^\mu$ can cross it no more than once.  Thus they define a continuous one-to-one map $E^+({\cal T})\to {\cal S}$.  Such a map, however, is possible only if ${\cal S}$ is itself compact.\footnote{This step uses the assumption that spacetime is connected, because otherwise ${\cal S}$ could contain multiple components, only some of which are noncompact.} Hence, with the assumptions (a)-(c) the spacetime cannot be future null-complete: at least one of the geodesics on $E^+({\cal T})$ must terminate at a singularity before reaching a conjugate point.
\end{proof}
 
\section{Including black holes}\label{inc}

We will now show that the Penrose theorem can be extended to allow for black hole formation.  Like the original Penrose theorem, our result can be applied in any spacetime dimension $D > 2$.  We shall use a generalized notion of black holes, to include dynamically evolving black brane-like solutions with arbitrary dimension and topology.\footnote{ In four dimensions, this would in principle allow black brane-like objects of dimension 1 and 2.  Note however, that in four dimensional general relativity, unlike in higher dimensions, there are no such stationary brane solutions in asymptotically flat spacetime \cite{Hawking2}.  A further constraint on 1-branes comes from the topological censorship theorem \cite{Friedman}, which in certain contexts requires the topologies of horizons to be simply connected \cite{Jacobson}.}  We will also include objects like the Einstein-Rosen bridge bounded by two or more disconnected horizons, although later we shall show that excluding such black holes leads to a stronger result.
 
Let us assume that there exist at least some future-infinite (timelike or null) worldlines, and thus a nonempty set of points at future infinity ${\cal I}^+$.  The union of the event horizons of all black holes can then be defined as
\beq
{\cal H}={\partial I}^-({\cal I}^+),
\label{calH}
\eeq
with each connected part of ${\cal H}$ representing an individual black hole horizon.  Similarly one can define past horizons using past infiniinfinityty ${\cal I}^-$.

In the proofs that follow, we will not be interested in the interior regions of the black holes.  Consequently, when we assume the existence of a trapped surface ${\cal T}$, we will need to require that it lie at least partly outside of ${\cal H}$.  More generally, we can define trapped surfaces so as to include compact $D - 2$ surfaces with a boundary $\partial {\cal T} \in {\cal H}$.\footnote{Note that in order for ${\cal T}$ to be compact, it must be a closed set in the sense of including its own boundary.}

Let us now discuss the types of singularities which can appear in the spacetime.  
We will define a singularity to be ``naked'' if a) it lies in $J^+(p)$ for some point $p$ but b) it also lies in $J^-({\cal I}^+)$, the region which can be seen by an observer outside of any horizons.  Criterion (a) excludes garden variety initial singularities, such as appear in the interior of a Schwarzschild white hole, or at the Big Bang of an FLRW model.

In the theorems that follow, we will assume that the spacetime is globally hyperbolic, i.e. that it admits a Cauchy surface.  We will also assume that there are no naked singularities which lie \emph{on} the black hole horizon itself (because of the use of $J^-$ in criterion (b), such singularities would still be naked, even though they are not ruled out by global hyperbolicity).  We shall now prove two theorems extending the Penrose's result.

{\bf Theorem 1:} Let ${\cal M}$ be a spacetime that obeys NCC, has a non-compact, connected Cauchy surface ${\cal S}$, and contains some black holes.  Let all singularities be in the interior of the event horizon ${\cal H}$, i.e. there are no naked singularities on or outside the horizon.  Suppose also that ${\cal M}$ contains a trapped surface ${\cal T}$ outside of the black hole horizons.  (In the case where a trapped surface is partly inside and partly outside of ${\cal H}$, we can without loss of generality consider ${\cal T}$ to include only the part which is outside.)  Then ${\cal T}$ must be completely surrounded by an event horizon.

Since event horizons can come into existence, whether or not ${\cal T}$ is completely surrounded by an event horizon might depend on the choice of time slice.  Therefore, we define ``completely surrounded'' more precisely to mean that there exists some edgeless achronal slice ${\cal F}$, such that ${\cal T}$ is contained in a compact subregion ${\cal R}$ of ${\cal F}$, and the boundary of ${\cal R}$ is part of ${\cal H}$.  This indicates that if one waits long enough, a black hole horizon will form outside of ${\cal T}$.  Moreover, because ${\cal F}$ is achronal, no causal curve which starts on ${\cal T}$ and moves to the future can avoid being enclosed by the horizon.  The same applies if you start on any compact spatial region whose edge is ${\cal T}$.

\begin{proof}
We again consider the future light cone of the trapped surface, $E^+({\cal T})$.  The null geodesics on $E^+({\cal T})$ converge and either reach their conjugate points outside of black hole horizons, or else cross one of the horizons.  (Inside of the horizon, a geodesic may either have a conjugate point or else terminate on a singularity in the interior.  Since there are no naked singularities, it cannot terminate on a singularity at the event horizon itself.)  The part of $E^+({\cal T})$ lying outside of black hole horizons is an achronal compact hypersurface, but now it can have edges at the intersections of $E^+({\cal T})$ with the horizons.

To remedy this situation, we construct a new surface ${\cal E}$ by joining $E^+({\cal T})$ with black hole horizons along the intersections, including only the part of $E^+({\cal T})$ which is to the past of ${\cal H}$ and vice versa (see Fig. \ref{surround}).  This new surface can also be defined as a boundary of a set,
\beq
{\cal E}=\partial [({\cal M} - I^+({\cal T})) \cap I^-({\cal I}^+)].
\eeq
It follows that ${\cal E}$ is edgeless (i.e. it has no boundary, except possibly at infinity or at a singularity).  Furthermore, the definitions (\ref{E+}), (\ref{calH}) imply that the surfaces $E^+({\cal T})$ and ${\cal H}$ are achronal.  In order to show that the combined surface ${\cal E}$ is also achronal, we have to make sure that no point on the relevant parts of ${\cal H}$ can be connected to the part of $E^+({\cal T})$ outside of horizons by a timelike curve.   This is indeed the case, since all future-directed timelike curves originating on ${\cal H}$ are contained entirely inside the horizons, and all future-directed timelike curves originating on $E^+({\cal T})$ are contained in $I^+({\cal T})$.

If some black holes never cross $E^+({\cal T})$, then ${\cal E}$ will contain multiple connected components.  In this case we will only be interested in the connected component of ${\cal E}$ which includes $E^+({\cal T})$.  Let us call this connected component ${\tilde{\cal E}}$, and let ${\tilde{\cal H}}$ be the part of ${\cal H}$ which is in ${\tilde{\cal E}}$.

Although ${\tilde{\cal E}}$ is edgeless, it cannot be compact.  For the existence of an edgeless, compact, achronal hypersurface is inconsistent with a non-compact Cauchy surface ${\cal S}$, as discussed in section \ref{pen}.  Therefore, ${\tilde{\cal E}}$ must have a ``boundary'' at infinity, or perhaps at an initial singularity, if ${\cal H}$ extends back that far in time.  

Nevertheless, $E^+({\cal T}) \cap I^-({\cal I}^+)$ is still compact, and its edge is contained in ${\tilde{\cal H}}$.  By making the identification\footnote{We remind the reader that ${\cal F}$ is an edgeless achronal slice which was introduced above in the definition of ``completely surrounded''.} ${\tilde{\cal E}} = {\cal F}$, we see that ${\cal T}$ is completely surrounded by the event horizon ${\tilde{\cal H}}$.  
\end{proof}

Apart from the ``bag of gold'' spacetime, another example of a situation where ${\cal T}$ is completely surrounded by the horizon is an effectively (1+1)-dimensional spacetime, where all but one of the spatial dimensions are compactified.  In this case, a sufficiently large black hole will always cut space into two disconnected pieces.  After the black hole originates, its horizon also splits up into two disconnected, spatially compact pieces.  A pair of black holes of this kind are capable of completely surrounding a trapped surface ${\cal T}$.  Note that in both examples black hole horizons divide ${\cal M}$ into causally disconnected parts.

We now turn to the second theorem, which makes a different assumption about the black holes.  Instead of assuming that they do not completely surround the event horizon, one can also get a contradiction by assuming that each black hole horizon originates, not at an initial singularity, but at some ``origin set'' ${\cal O} \in {\cal M}$, and that after $\cal{O}$ each black hole contains just one connected component.

{\bf Theorem 2:}  Let ${\cal M}$ be a spacetime that obeys NCC, has a non-compact, connected Cauchy surface ${\cal S}$, and contains some black holes.  Let all singularities be in the interior of the event horizon ${\cal H}$, i.e. there are no naked singularities on or outside the horizon.  Suppose also that ${\cal M}$ contains a trapped surface ${\cal T}$ which is outside of the black holes.  (Again, if a trapped surface is partly inside and partly outside, we only consider the piece of ${\cal T}$ which is outside.)  Then either (a) ${\cal H}$ extends all the way back in time to $\cal I^-$ or to an initial singularity, or else (b) there exist black holes whose horizons contain multiple connected components (i.e. some connected component of ${\cal H}$ breaks up into multiple connected components of ${\cal H-O}$, as in the case of the Einstein-Rosen bridge discussed above.)

\begin{proof}
To prove this result, we must describe in greater detail the structure of the horizon.  Each of the horizon generators in ${\tilde{\cal H}}$ can be labelled with an affine parameter $\lambda$, which we take to be increasing towards the future.  Note that generators can enter a future horizon, but they can never leave it \cite{HawkingEllis}.  If (a) is false, then each generator must enter the horizon at a finite value of the affine parameter.  The space of points in ${\cal M}$ at which horizon generators enter the horizon is the origin set $\cal{O}$.  $\cal{O}$ is an acausal (i.e. strictly spacelike) set, which can contain pieces whose dimensionality ranges from $0$ to $D-2$, $D$ being the spacetime dimension.  We now divide the null rays generating ${\tilde{\cal H}}$ into two classes: 
\begin{enumerate}[i.]
\item those which never intersect $E^+({\cal T})$, and 

\item those which eventually intersect $E^+({\cal T})$ at some finite affine time $\lambda = \lambda_*$.
\end{enumerate}
If we assume that all generators of ${\tilde{\cal H}}$ are intersected by the surface $E^+({\cal T})$, as would happen if ${\tilde {\cal H}}$ consists entirely of isolated black hole horizons, then class (i) is empty and we can restrict attention to class (ii).  But then ${\tilde{\cal H}}$ is compact, and the part of $E^+({\cal T})$ outside of black holes is also compact; hence ${\tilde{\cal E}}$ is compact.  We thus conclude that ${\tilde{\cal E}}$ is an edgeless, compact, achronal hypersurface.  The existence of such a hypersurface is inconsistent with a non-compact Cauchy surface ${\cal S}$, so we obtain a contradiction.

The alternative is to suppose that class (i) is \emph{not} empty.  In this case ${\tilde{\cal E}}$ is noncompact, since each generator in class (i) extends infinitely far to the future.  Although ${\tilde{\cal E}}$ is edgeless, it still has a ``boundary at infinity'', which is part of ${\cal I}^+$.  Hence ${\tilde{\cal E}}$ is not a \emph{complete} achronal surface, i.e. not every inextendible causal curve will intersect it.

Note that because $E^+({\cal T}) \cap I^-({\cal I}^+)$ is compact, its edge must also be compact.  This edge is the locus of points for which $\lambda = \lambda_*$.  Since this edge is shared with generators in class (ii), it follows that the class (ii) piece of ${\tilde{\cal H}}$ is compactly generated.

Next we have to consider the way in which the class (ii) generators may be joined to the class (i) part of ${\tilde{\cal H}}$.  These two parts of the horizon cannot be glued together along some subset $G$ of horizon generators, because that would require that for generators $g^\prime$ near $G$ on the class (ii) side, that $\lambda_*(g^\prime) \to +\infty$ as $g^\prime \to g\in G$.  But then the boundary at $\lambda = \lambda_*$ would be noncompact, which is disallowed.

In other words, if $E^+({\cal T})$ intersects any generator of ${\tilde{\cal H}}$, it must also intersect every other generator in the same connected component of $\tilde{{\cal H}} - {\cal O}$.  This means that whenever $E^+({\cal T})$ hits a black hole whose horizon has just one connected component after it collapses, then it completely wraps around the black hole.  It cannot intersect some of the generators but not others.

Hence, the class (i) generators must be joined to the class (ii) generators at ${\cal O}$, where they enter the horizon.  Thus the set ${\cal O}$ will generally divide ${\tilde{\cal H}}$ into a number of connected components, some of which are type (i) and others type (ii).  Thus, on the assumption that (a) is false, we have shown that (b) is true.  
\end{proof}

We further note that if a simply connected spacetime has a black hole with two or more connected components (i.e. if it satisfies (b)), then after the formation of the black hole, the exterior of the horizon will be divided into multiple causally disconnected regions.  That is, if two components ${\tilde{\cal H}}', {\tilde{\cal H}}'' \subset {\tilde{\cal H}}-{\cal O}$ join one another at ${\cal O}' \subset {\cal O}$, then the parts of ${\cal M}$ facing different components of the horizon are causally disconnected from one another.  More precisely, the region
${\cal M} - (J^+({\cal H}) \cup  J^-({\cal O}'))$
must split into multiple components, none of whose points are connected by a causal curve in ${\cal M}$.
This is because they are connected, but not causally connected, through the black hole horizon.  Because the spacetime is simply connected, they cannot also be causally connected through some path which does not cross through the black hole horizon.\footnote{If spacetime is not simply connected, we can pass to the ``universal cover'' of the spacetime, which is simply connected.  This universal cover must then have causally disconnected regions.}

\section{Summary and discussion}

The two theorems we proved in Section III are similar to Penrose's singularity theorem \cite{Penrose}.  As in Penrose's theorem, we assume that the spacetime obeys the null convergence condition, has a non-compact connected Cauchy surface, and includes a trapped surface.  Assuming in addition that the trapped surface lies (at least partly) outside the black hole horizons and is not completely surrounded by horizons, we proved in Theorem 1 that the spacetime must contain naked singularities on the horizon.  
Thus, unlike the Penrose's original theorem, our theorem provides some information about the nature and the location of the singularities.

The examples we discussed to illustrate how a trapped surface ${\cal T}$ can be surrounded by a horizon suggested that this can only happen when ${\cal T}$ is located in a spatially compact region which is causally disconnected from the rest of spacetime by black hole horizons.   This led us to Theorem 2, which shows that if the black hole horizon surrounding ${\cal T}$ originates at some finite time (rather than at an initial singularity or past infinity), then it must have multiple connected components.  But if the horizon does have multiple connected components, then on a simply connected spacetime, once the black hole appears it must divide the spacetime into causally disconnected pieces.

Another mechanism that can generate singularities out of non-singular initial data is the formation of negative-energy, Anti-de Sitter (AdS) bubbles.  The interiors of such bubbles have the geometry of open $(k=-1)$ FLRW universes, and the negative energy density of the vacuum causes them to contract and collapse to a big crunch \cite{ColemanDeLuccia}.  The resulting singularity is contained within the future light cone of the nucleation center (the `bubble cone').  Since these count as black hole horizons in \eqref{calH}, Theorems 1 and 2 already cover this case.

To illustrate how these theorems can be used to rule out cosmological models, we consider the Carroll-Chen (CC) scenario as an example (see also \cite{AV}).  The starting point of this scenario is that some typical (or ``generic''), regular boundary conditions are specified on an infinite spacelike Cauchy surface ${\cal S}$.  This initial state is then evolved both to the past and to the future of ${\cal S}$.  Assuming that the lowest-energy (true) vacuum has a positive energy density, CC argue that (i) the universe approaches an empty true-vacuum de Sitter configuration in both time directions, and that (ii) the resulting spacetime is non-singular, apart from isolated black holes that may be formed in the future and/or in the past of ${\cal S}$ (see Fig.~1).

If (i) is true, then the asymptotic de Sitter region to the past of ${\cal S}$ necessarily contains trapped surfaces.  For example, any spherical surface of radius greater than the de Sitter horizon is trapped.  Then, assuming that NCC is satisfied, it follows from Theorems 1 and 2 that (ii) must be false---there are either naked singularities, or else the ``black hole'' singularities are sufficiently extensive to divide the spacetime into multiple compact regions.

In order to save the CC scenario, one must change one of the assumptions.  For example, one could suppose that the typical state on ${\cal S}$, rather than evolving to an isolated black hole surrounded by inflating regions, instead evolves mostly to singular configurations, with isolated regions which inflate.  In that case, the isolated regions will be surrounded by one, topologically complicated ``black hole'', and neither Theorem 1 nor Theorem 2 will apply.

Alternatively, one could suppose that the universe actually started with a spatially compact Cauchy surface ${\cal S}$.  In this case, there would be no guarantee that any inflating regions would exist in a typical state.  If ${\cal S}$ were sufficiently large, it is conceivable that (for some definition of ``typical'' states) there would be a high probability of containing at least one inflating region.  However, some arguments against this possibility have been given in \cite{AV}.

\subsection*{Acknowledgements}

This work was supported in part by the National Science Foundation grant PHY-1213888 and by the Templeton Foundation (A.V.), and by the Simons Foundation and the National Science Foundation grant PHY-1205500 (A.W.).

\end{document}